\documentclass[useAMS,usenatbib]{mn2e}

\usepackage{amsmath}
\usepackage{amssymb}
\usepackage{graphicx}
\usepackage{dcolumn}
\usepackage{bm}
\usepackage{ulem}

\title[Liquid-phase epitaxy of neutron star crusts]
{Liquid-phase epitaxy of neutron star crusts and white dwarf cores}
\author[D. A. Baiko]{D. A. Baiko\thanks{E-mail:baiko@astro.ioffe.ru} \\
Ioffe Institute, Politekhnicheskaya 26, 194021 Saint Petersburg, Russia}

\begin{document}

\date{Accepted; Received ; in original form}

\pagerange{\pageref{firstpage}--\pageref{lastpage}} \pubyear{2014}

\maketitle

\label{firstpage}

\begin{abstract}
Near-equilibrium bottom-up crystallization of fully-ionized neutron star
crusts or white dwarf cores is considered. We argue that this process is 
similar to liquid-phase epitaxial (i.e. preserving order 
of previous layers) crystal growth or crystal pulling from 
melt in Earth laboratories whereby lateral positions of newly crystallizing ions are anchored
by already solidified layers. Their vertical positions are set by
charge neutrality. Consequently, interplane spacing of a growing 
crystal either gradually
increases, tracing $n_\mathrm{e}$ decrease, as the crystallization front
moves away from the stellar center, or decreases, tracing decrease of 
$\langle Z \rangle$, when the crystallization front crosses a boundary
between layers of different compositions. This results in a formation of
stretched Coulomb crystals, in contrast to the standard 
assumption of cubic crystal formation, 
which is based on energetics arguments but does not take into account 
growth kinetics. 
Overstretched crystals break, which
limits the vertical sizes of growing crystallites. We study breaking 
shear strain and effective shear modulus of stretched matter and discuss 
possibility of macrocrystallite formation. The latter has interesting 
astrophysical implications, for instance, appearance of weak 
crustal layers, whose strength may increase by a few orders of 
magnitude upon breaking and refreezing at a late-time event. We also 
analyze interaction of adjacent Coulomb crystals, having different ion 
compositions, and estimate the strength of such interfaces.   
\end{abstract}

\begin{keywords}
dense matter -- equation of state -- stars: neutron -- white dwarfs.
\end{keywords}



\section{Introduction}
Neutron stars (NS) and white dwarfs (WD) are magnificent astrophysical
objects renowned for unparalleled 
breadth and significance of observational 
manifestations as well as for absolutely extreme physical conditions 
in their interior \citep*[e.g.][and references therein]{FB08,WK08,A+10,
K10,KK15,MPM15,OF16,KB17,C+19,SBT22}. These stars, with masses in 
excess of 2 M$_\odot$ for NS and up to about 1.4 M$_\odot$ for WD, 
with thermonuclear reactions turned off, are kept from gravitational collapse 
by pressure of degenerate fermions, neutrons and electrons, 
respectively \citep*[e.g.][]{HPY07}. Their central densities are expected to be 
$\sim 10^{15}$ g/cc for NS and up to $\sim 10^{10}$ g/cc for WD. At 
densities $\rho$ below the nuclear saturation density, 
$\rho_0 = 2.8 \times 10^{14}$ g/cc, i.e. in the outer layers 
of NS called crust and in the entire WD interior, structure of matter 
remotely resembles that of Earth metals. There are fully-ionized ions,
i.e. atomic nuclei, which, in deeper layers of NS, become 
neutron-rich, and strongly degenerate nearly uniform electron gas. The 
electron gas is ultrarelativistic at $\rho \gg 10^6$ g/cc. In NS
at $\rho > \rho_\mathrm{d} = 4.3 \times 10^{11}$ g/cc, there is 
also Fermi-liquid of neutrons dripped from the nuclei. In the very outer
layers at $\rho \lesssim 10^4$ g/cc (depending on ion sort),
conditions of full ionization and electron degeneracy gradually 
cease to exist.

Ion composition of such dense matter is a complicated topic 
\citep[e.g.][and references therein]{BPY21,SGC23}. In 
general, the composition varies strongly with density. For WD, one typically
considers less massive stars made of helium, stars of intermediate mass
which contain a C/O mixture with a pronounced transition from 
deeper O-enriched regions to C-enriched matter in the outer core, and 
finally, the most massive stars made of an O/Ne mixture with 
a possibility of Mg, Si etc. Also there are smaller admixtures of 
various other elements and isotopes. In NS, though lighter elements
may be present at relatively low densities, one typically 
considers heavier nuclei, for instance, iron (which represents matter ground 
state at the lowest densities) and many other. The composition in this 
case is calculated by minimizing thermodynamic potential of matter or
by following a nuclear reaction network, taking place in a freshly 
accreted fuel. The end result of such calculations is a sequence of 
shells, each corresponding to a specific range of mass densities with
composition dominated by a particular atomic nucleus. Such a structure
is sometimes referred to as the ``onion'' structure of the NS crust.  
Predicted isotopes oftentimes have different charge-to-mass ratios and 
tend to gravitationally separate making the ``onion'' structure even 
more pronounced.

As an NS or a WD cools, their internal temperature drops below local 
crystallization temperature, which results in a gradual freezing
of matter from the deeper layers up. Formed crystals, which are called 
Coulomb crystals to emphasize long-range nature and simplicity of 
forces, binding them together, are typically assumed to have (the most 
tightly-bound) body-centred cubic (bcc) lattice. 
In what follows, we shall neglect electron screening, i.e. slight 
deviations of electron density from uniformity. This is a well 
justified approximation in compact degenerate stars at not too low densities, 
i.e. down to densities, where full-ionization approximation starts 
breaking down \citep[e.g.][]{HPY07}. Hence in our approach, electrons will be treated as an
ideal constant density background. Besides that, electrons are responsible 
for thermal and electric conductivities of matter, which are believed
to be very high \citep[e.g.][]{PPP15} and thus provide for an exclusively 
uniform thermal environment.      

In this paper, we take a fresh look at the bottom-up crystallization process in
dense matter of compact stars and critically reappraise the bcc 
lattice assumption, breaking strain, and shear modulus for these objects.
Also, using methods of lattice dynamics, we study an idealized model of an 
interface between layers with different compositions, e.g. shells of the 
``onion'' structure in NS crust. In particular, we analyze interaction 
between different layers and try to answer the question as to the effect 
of such interfaces on the overall crustal strength.
A related problem of spherical Coulomb crystal energy, which is 
relevant for dusty plasma crystallization and ion crystallization in 
traps, has been investigated e.g. by \citet[and references therein]{HA91}.

In section \ref{GF}, we deduce a formula for the interaction energy of 
a charge with a Coulomb half-crystal, consisting of ordered planar  
layers of identical ions immersed into charge-neutralizing electron background. 
In section \ref{ME}, this formula is manipulated
to produce an alternative expression for the Madelung energy, which
reproduces previously known result for the bcc lattice. 
In section \ref{NSF}, we consider interaction of two Coulomb crystals 
composed of ions of two different sorts, propose a simplified model 
of the ``onion'' structure boundary, and analyze its breaking 
properties. In section \ref{EPI}, which is central for the whole work, 
we establish the possibility of epitaxial crystal growth in dense 
matter of degenerate stars and predict crystal stretching effect, 
associated with it.   
In section \ref{BSSM}, breaking properties of stretched matter are 
studied quantitatively. In particular, we consider NS crust or WD core 
composed of crystallites with random orientations of crystallographic 
planes with respect to the stretch direction (polycrystalline model). By contrast, in section 
\ref{SNGL}, we address the possibility of large-scale 
crystallite formation and possible relations between their properties and 
astrophysical phenomena (macrocrystallite model). In section \ref{ESM}, shear modulus of 
stretched matter is evaluated. Finally, in section \ref{OPE}, we 
list a few other effects, which are expected to accompany elongation and 
contraction of Coulomb crystals in NS and WD interior.

\section{Interface of two crystals}
\subsection{Interaction of a charge with a crystal}
\label{GF}
We would like to begin by studying the interaction energy of a charge 
$Q$ with a plane, containing a 2D lattice of ions with charges $Ze$ ($e$ is the positron charge), 
spanned by basis lattice vectors $\bm{a}_1$ and $\bm{a}_2$, and a slab
of neutralizing electrons of constant density $n_{\rm e}$, enveloping symmetrically the 
ion lattice. If $Q$ is outside of the 
electron slab, the latter can be shrunk to the same plane with surface 
density $\sigma = n_{\rm e} \eta$, where $\eta$ is the slab width.
    
Let us pick an ion of the lattice and define its position as the origin.
Let the position vector of the charge $Q$ be 
$\bm{d}=\bm{d}_\perp + \bm{d}_\parallel$. In this case, index $\perp$ 
indicates vectors perpendicular to the plane whereas index $\parallel$ 
marks in-plane vectors. Then the interaction energy $U_1$ is given by
\begin{equation}
    \frac{U_1}{QZe} = \sum_{\nu,\mu=-\infty}^{+\infty} \frac{1}{R} -
    \frac{\sigma}{Z} 
    \int_{\rm 2D} \frac {{\rm d} \bm{r}}{|\bm{r}-\bm{d}|}~,
\label{U1def}
\end{equation}
where $\bm{R} = \bm{R}_\perp+\bm{R}_\parallel$ is the difference between
an ion position vector and the charge $Q$ position vector, 
$\bm{R}_\perp = - \bm{d}_\perp$,
$\bm{R}_\parallel = \nu \bm{a}_1 + \mu \bm{a}_2 - \bm{d}_\parallel$. 
Furthermore,
\begin{eqnarray}
   \frac{1}{R}  &=& \frac{2}{\sqrt{\upi}} \int_0^{+\infty} {\rm d} \rho
   \mathrm{e}^{-\rho^2(R^2_\perp + R^2_\parallel)} \\
   &=&
   \frac{2}{\sqrt{\upi}} \int_0^{+\infty} {\rm d}\rho 
   \mathrm{e}^{-\rho^2 R^2_\perp}
   \frac{\upi}{\rho^2} \int_{\rm 2D} \frac{{\rm d}\bm{q}}{(2\upi)^2}
   \mathrm{e}^{-q^2/4\rho^2 + \mathrm{i}\bm{q}\bm{R}_\parallel}~,
\nonumber 
\end{eqnarray}
and an analogous representation can be written for the second term in
equation (\ref{U1def}). Using the fact that
\begin{eqnarray}
     \sum_{\nu,\mu} \mathrm{e}^{\mathrm{i}\bm{q}(\nu \bm{a}_1 
     + \mu \bm{a}_2)} &=&
     (2 \upi)^2 \frac{\sigma}{Z} \sum_{\bm{G}} \delta(\bm{q}-\bm{G}) \\
       \int_{\rm 2D} {\rm d}\bm{r} \mathrm{e}^{\mathrm{i}\bm{q}\bm{r}} 
       &=& (2\upi)^2 
       \delta(\bm{q})~,   
\end{eqnarray}
where $\bm{G}$ are 2D reciprocal lattice vectors for direct vectors
spanned by $\bm{a}_1$ and $\bm{a}_2$, we deduce 
\begin{equation}
      \frac{U_1}{QZe} =  \frac{2}{\sqrt{\upi}} \int_0^{+\infty} 
      {\rm d}\rho \mathrm{e}^{-\rho^2 R^2_\perp} 
      \frac{\upi \sigma}{\rho^2 Z}
      \sum_{\bm{G}\ne0} \mathrm{e}^{-G^2/4\rho^2 
      + \mathrm{i}\bm{G}\bm{r}_\parallel}~.
\label{U1au}
\end{equation}
In this case, $\bm{r}_\parallel = - \bm{d}_\parallel$. 
The $\rho$-integral can be evaluated explicitly with the result
\begin{equation}
   U_1 = 2\upi Qe \sigma \sum_{\bm{G}\ne0}
   \frac{1}{G} \, 
   \mathrm{e}^{-G R_\perp+ \mathrm{i}\bm{G}\bm{r}_\parallel}~.
\label{U1}
\end{equation}

Suppose now that there is not one ion plane but half a space filled 
with such planes separated by distance $\eta$. Ion positions in the 
$\kappa$-th plane ($\kappa=0,1,2,\ldots$) are obtained from ion 
positions in the original plane by adding an out 
of plane vector $\kappa\bm{a}_3$ 
($\bm{a}_3=\bm{a}_{3\perp}+\bm{a}_{3\parallel}$, 
$|\bm{a}_{3\perp}|=\eta$). We note in passing that any simple 3D lattice 
can be specified in this way.  

Interaction energy of the charge $Q$, located 
outside of the half-space,
with each of these planes is given by the same equation (\ref{U1})
with $\bm{R}_\perp=\kappa \bm{a}_{3\perp} - \bm{d}_\perp$ and 
$\bm{r}_\parallel=\kappa \bm{a}_{3\parallel} - \bm{d}_\parallel$. 
Moreover, one can sum over $\kappa$ and find interaction energy $U$ of 
the charge with the whole half-space as
\begin{equation}
     U = 2 \upi Qe \sigma \sum_{\bm{G}\ne0}
     \frac{\mathrm{e}^{-G|\bm{d}_\perp|-\mathrm{i} \bm{G} 
     \bm{d}_\parallel}}
     {G\left(1-\mathrm{e}^{-G \eta + \mathrm{i}\bm{G} 
     \bm{a}_{3\parallel}}\right)}~.
\label{U}
\end{equation}

\subsection{Madelung energy of a crystal}
\label{ME}
As a first application of this formula,
we shall reproduce the Madelung
energy of a 3D ion lattice based on it. We suppose, that 
$Q=Ze$, $\bm{d}=-\bm{a}_3$, and, moreover, that the charge $Q$ 
belongs to a plane of identical charges, adjacent to the filled 
half-space and positioned and oriented with respect to the original 
plane and the half-space as it would be in an ideal 3D crystal. Let 
this plane be also immersed into its own electron slab of the same 
density $n_{\rm e}$ and thickness $\eta$. Then, it is necessary to find 
the interaction energy of $Q$ with other charges in its plane. Using 
Ewald transformation, we can write
\begin{eqnarray}
       && \frac{U_0}{Z^2e^2} = \sum_{\bm{R}\ne 0} \frac{1}{R} - 
       \frac{\sigma}{Z} \int_{\rm 2D} \frac{{\rm d}\bm{r}}{r}
\nonumber \\
       &=& \frac{2}{\sqrt{\upi}} \int_{\cal A}^{+\infty} {\rm d}\rho 
       \left( \sum_{\bm{R}\ne 0} \mathrm{e}^{-\rho^2 R^2} 
       - \frac{\sigma}{Z} 
       \int_{\rm 2D} {\rm d}\bm{r} \mathrm{e}^{-\rho^2 r^2} \right)
\nonumber \\
       &+&  \frac{2}{\sqrt{\upi}} \int_0^{\cal A} \frac{\upi{\rm d}\rho}{\rho^2}
       \int_{\rm 2D} \frac{{\rm d}\bm{q}}{(2\upi)^2} 
       \mathrm{e}^{-q^2/4\rho^2}
\nonumber \\
       &\times&
       \left(\sum_{\bm{R}} \mathrm{e}^{\mathrm{i}\bm{q}\bm{R}} - 1 
       - \frac{\sigma}{Z}
       \int_{\rm 2D} {\rm d}\bm{r} \mathrm{e}^{\mathrm{i}\bm{q}\bm{r}} 
       \right)
       = \sum_{\bm{R}\ne0} \frac{1}{R} {\rm erfc}({\cal A}R) 
\nonumber \\
          &+& \frac{2\upi\sigma}{Z} \sum_{\bm{G}\ne0} \frac{1}{G} 
          {\rm erfc} \left(\frac{G}{2{\cal A}}\right)
          - \frac{2{\cal A}}{\sqrt{\upi}} -\frac{2\sqrt{\upi}\sigma}{{\cal A}Z}~,
\label{U0}
\end{eqnarray}
where ${\cal A}>0$ is an arbitrary real number, 
$\bm{R} = \nu \bm{a}_1 + \mu \bm{a}_2$,  
$\sum_{\bm{R}} = \sum_{\nu,\mu}$,
and $\sum_{\bm{G}}$ is the same as in the equation (\ref{U1au}).  

Madelung energy is the crystal electrostatic energy {\it per ion}. 
Only half of $U_0$ thus contributes to the Madelung energy. This is 
pretty obvious
for the $1/R$ term, but the situation with the ion-background term
in the first line of equation (\ref{U0})
is slightly more subtle. For Madelung energy, it should be replaced by 
two other terms, an ion-background contribution equal to
\begin{equation}
      -Ze \cdot e n_{\rm e} 
      \int_{-\eta/2}^{+\eta/2} {\rm d}z \int_0^{\tilde{r}} 
      \frac{2\upi r {\rm d}r}{\sqrt{r^2+z^2}}~,
\end{equation}
and a background-background term equal to
\begin{equation}
      + \frac{1}{2}\,e n_{\rm e} S \cdot e n_{\rm e} 
      \int_{-\eta/2}^{+\eta/2} {\rm d}z^\prime 
      \int_{-\eta/2}^{+\eta/2} {\rm d}z \int_0^{\tilde{r}} 
      \frac{2\upi r {\rm d}r}{\sqrt{r^2+(z-z^\prime)^2}}~,
\end{equation}
where $\tilde{r} \to \infty$, and the area $S=Z/n_{\rm e} \eta$. Such a
replacement results in a correction of 
$U_{\rm bg}= \upi Z e^2 n_{\rm e} \eta^2/6$.

For the $U$ contribution, equation (\ref{U}), factor $1/2$ is not needed 
because $U$ 
describes interaction with a half-space, but the second half-space
produces equal contribution. It is easy to verify, that, for this 
term, there are no corrections associated with treatment of the 
background. Thus
\begin{equation}
       U+\frac{1}{2}U_0+U_{\rm bg} = 
       \zeta U_{\rm C} \equiv \zeta \frac{Z^2 e^2}{a_{\rm i}}~,
\label{zeta}
\end{equation}
where $\zeta$ is the Madelung constant, 
$a_{\rm i}=(4\upi n_{\rm i}/3)^{-1/3}$ is the ion sphere
radius, and $n_{\rm i} = n_{\rm e}/Z$ is the ion 3D density. 
For bcc lattice described e.g. by
vectors $\bm{a}_1 = (1,0,0)\,a_{\rm l}$, 
$\bm{a}_2 = (0,1,0)\,a_{\rm l}$, 
$\bm{a}_3 = (-1/2, -1/2, -1/2)\,a_{\rm l}$, where $a_{\rm l}=2\eta$ is 
the cube edge ($n_{\rm i} a^3_{\rm l}=2$), we obtain from equation 
(\ref{zeta}) $\zeta=-0.895\,929\,255\,682$ in full 
agreement with \citet{BPY01}.

\subsection{Interaction of two crystals}
\label{NSF}
As another application of equation (\ref{U}), we shall consider a simple 
model of an interface in the ``onion'' structure
predicted to form in NS crust. Accordingly, we 
suppose that there are two semi-infinite one-component crystals composed 
of ions, one with charges $Z_1 e$, the other with charges $Z_2 e$, 
$Z_2>Z_1$, touching along some plane. This plane can belong to different 
crystallographic families in the two crystals. Without 
loss of generality, we shall assume that the $Z_2$-phase is ``below'' 
the $Z_1$-phase. The $Z_2$-phase can be identified with 
the half-space discussed in section \ref{GF}, whereas charges $Z_1 e$ can be 
identified with $Q$. The $Z_1$-lattice is then spanned by basis vectors 
$\bm{b}_1$ and $\bm{b}_2$, belonging to the same plane as the vectors
$\bm{a}_1$ and $\bm{a}_2$, and by an out of plane vector $\bm{b}_3$.
If we pick a $Z_1$-ion in the bottom plane of the $Z_1$-structure and 
denote its position as $\bm{d}$ (the origin is still at one of the 
$Z_2$-ions of the top plane of the $Z_2$-half-space), positions of all 
the other $Z_1$-ions in this plane are given by 
$\bm{d}+n\bm{b}_1+m\bm{b}_2$. Ion positions in the $k$-th plane are 
obtained by adding vector $k\bm{b}_3$. In this case, 
$n,m=0,\pm 1,\pm 2,\ldots$, $k=0,1,2,\ldots$.

\begin{figure*}
\begin{center}
\leavevmode
\scalebox{1}[-1]{
\includegraphics[bb=163 138 1221 490,width=175mm,clip]{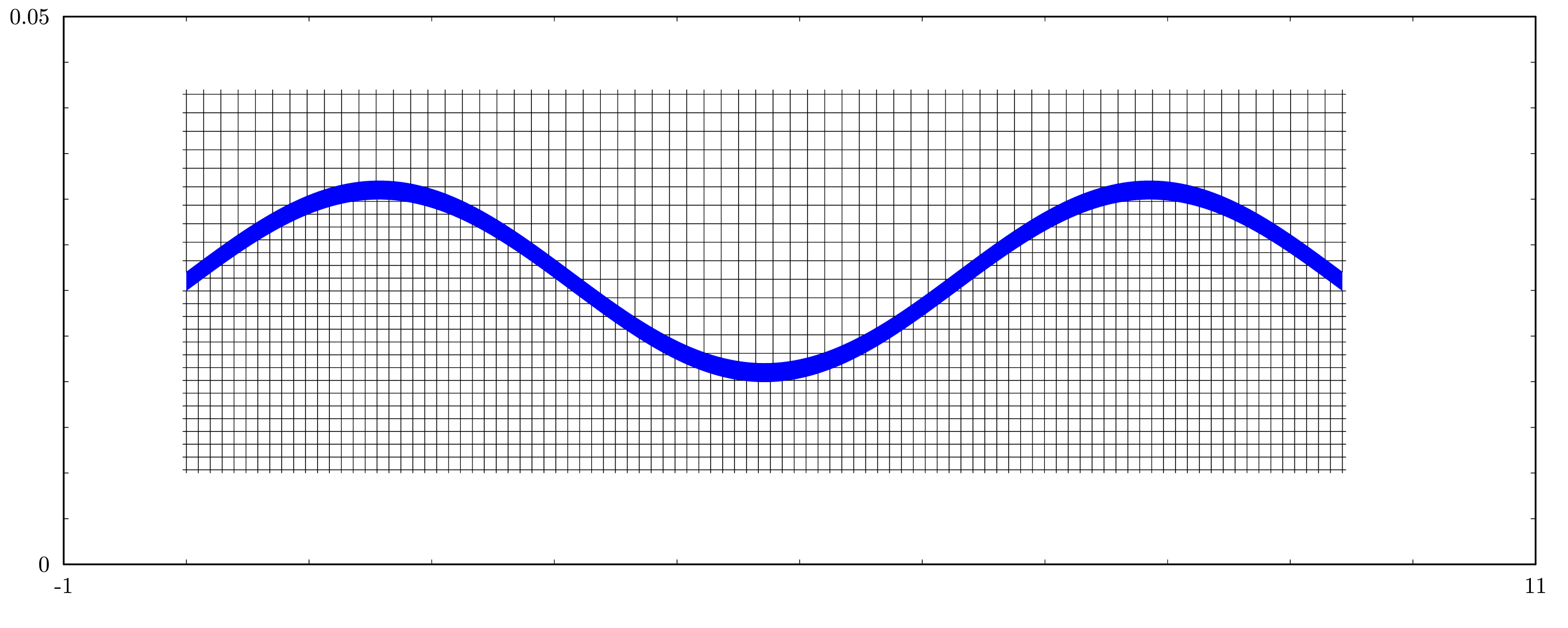}
}
\end{center}
\vspace{-0.2cm}
\caption[ ]{A schematic of the interface of two incommensurate crystals. 
Shaded blue region is the quasi-crystal zone. Lattice constants for 
both crystals are greatly exaggerated in comparison with the typical 
``wavelength'' of the interface roughness.}
\label{waves}
\end{figure*}

It is crucial for our argument, that the phases have essentially the 
same\footnote{It is easy to take into account a small jump of electron 
density between $Z_1$- and $Z_2$-phases due to ion Madelung pressure, 
but we shall not do that, as it does not modify our qualitative 
conclusions.} electron density $n_{\rm e}$, which, in the astrophysical 
context, is predominantly responsible for pressure and hydrostatic 
equilibrium. Then, if $Z_1$-lattice is also bcc, its cube edge 
$b_{\rm l} = (Z_1/Z_2)^{1/3} a_{\rm l}$. The interplane spacing (or the 
electron slab thickness) is $h=|\bm{b}_{3\perp}|$, and the distance 
from the top plane of the $Z_2$-crystal to the bottom plane of the 
$Z_1$-crystal is $|\bm{d}_\perp| = 0.5(\eta+h)$. If both cubes are 
aligned with the boundary plane, i.e. this plane is a $\{100\}$ plane 
for both crystals (more on this notation in section \ref{BSSM}), then
$h=b_{\rm l}/2=(Z_1/Z_2)^{1/3} a_{\rm l}/2 = (Z_1/Z_2)^{1/3} \eta$.

The interaction energy of each $Z_1$-ion with the $Z_2$-lattice
is given by equation (\ref{U}), in which one has to replace 
$|\bm{d}_\perp|$ by $|\bm{d}_\perp|+k|\bm{b}_{3\perp}|$
and $\bm{d}_\parallel$ by 
$\bm{d}_\parallel+n\bm{b}_1+m\bm{b}_2+k\bm{b}_{3\parallel}$. The 
sum over $Z_1$-ions yields the interaction energy\footnote{Obviously, 
electron background of the $Z_1$-crystal does not contribute, as it 
would contain integrals of equation (\ref{U}) over $\bm{d}_\parallel$, 
whereas $\bm{G}\ne 0$.} between the two crystals. The main
question then becomes, whether the interaction energy depends on 
$\bm{d}_\parallel$, i.e. on the $Z_1$-lattice lateral\footnote{Given 
lattice types and orientations, their vertical separation is set by 
$n_{\rm e}$.} position with respect to the $Z_2$-lattice. 

Summation of the factor 
$\exp{[-i \bm{G} (\bm{d}_\parallel+n\bm{b}_1+m\bm{b}_2
+k\bm{b}_{3\parallel})]}$ in equation (\ref{U}) over $n$ and $m$ 
restricts the sum over $\bm{G}$ in equation (\ref{U}) only to those 
$\{\bm{G}^\prime\} \subset \{\bm{G}\}$, which are reciprocal vectors 
simultaneously for direct vectors 
generated by $\bm{a}_1$ and $\bm{a}_2$ and by $\bm{b}_1$ and $\bm{b}_2$.   
For there to be any such reciprocal vectors, vectors $\bm{b}_1$ and $\bm{b}_2$ must be linear combinations
of $\bm{a}_1$ and $\bm{a}_2$ with rational coefficients $t_{ij}$: 
$\bm{b}_i = \sum_j t_{ij} \bm{a}_j$, $i,j=1,2$. Suppose, for brevity,
that vectors $\bm{a}_1$ and $\bm{a}_2$ are those given in the 
end of section \ref{ME}. Then the distance between $Z_1$-ions 
separated by the vector $\bm{b}_1$ is 
$a_{\rm l} \sqrt{t^2_{11} + t^2_{12}}$. On the other hand, the distance
between any two ions in a bcc lattice with the cube edge $b_{\rm l}$ is
$b_{\rm l} \sqrt{l/4}$ with $l$ integer. Equating and squaring, we
observe that for there to be any 
$\{\bm{G}^\prime\} \subset \{\bm{G}\}$,  
$(b_{\rm l}/a_{\rm l})^2 = (Z_1/Z_2)^{2/3}$ must be rational, which is 
satisfied for very rare $Z_1/Z_2$. For other charge ratios, 
there are no appropriate 
$\bm{G}^\prime$, and the interaction energy between $Z_1$- 
and $Z_2$-half-spaces is zero. 

Thus, the system is insensitive to their 
mutual orientation and lateral position (vertical separation 
is set by $n_{\rm e}$), so that they are free to move or rotate 
with respect to each other without any static friction 
(dynamic friction will be non-zero e.g. due to electron viscosity). We 
shall refer to such lattices as {\it incommensurate}.

In the same fashion, one can easily study a situation, in which one of 
the crystals (or both) is stressed in such a way that it is no 
longer cubic, but is commensurate with the lateral lattice spacing 
of the other crystal. Consequently, the set of $\bm{G}^\prime$ will not 
be empty, and the interaction energy will not be zero. Moreover, 
summation over $k$ can be performed then in a closed form with the 
result:
\begin{eqnarray}
     \hspace{-0.3cm} && U_{\rm int} = 2\upi Z_1 e^2 \sigma 
\nonumber \\ 
     \hspace{-0.3cm} && \times \sum_{\bm{G}^\prime\ne0}
     \frac{\mathrm{e}^{-G^\prime|\bm{d}_\perp|-\mathrm{i} \bm{G}^\prime 
     \bm{d}_\parallel}}
     {G^\prime \left(1-\mathrm{e}^{-G^\prime \eta 
     + \mathrm{i}\bm{G}^\prime 
     \bm{a}_{3\parallel}}\right)
     \left(1-\mathrm{e}^{-G^\prime h - \mathrm{i}\bm{G}^\prime 
     \bm{b}_{3\parallel}} \right)}~,  
\label{Uint}
\end{eqnarray}
where $U_{\rm int}$ is the total interaction energy of the crystals per 
each $Z_1$-ion of the bottom plane. This expression does depend on 
$\bm{d}_\parallel$.

The model of crystal interaction developed in this section is however 
too simplistic for two reasons. First is the fact, that, in the bottom 
plane of the $Z_1$-lattice as well as in the top plane of the 
$Z_2$-lattice, there is a complex set of potential wells, produced by 
both half-spaces (cf. section \ref{EPI}). Hence, it would be more realistic to assume 
the presence of at least one intermediate plane with a mixed composition between 
the two half-lattices. Using equation (\ref{U}) twice, one could 
generate a set of potential wells in this layer due to both half-spaces,
which would have an aperiodic 
quasi-crystal arrangement and different depths. It would be possible to
maintain overall charge neutrality by counting these wells, specifying 
the relative fractions of $Z_1$- and $Z_2$-ions, and adjusting the 
intermediate electron slab thickness accordingly. 

Our preliminary study indicates that the static friction in this model depends 
on ion placement in the wells, 
but, in any case, is orders of magnitude lower than that 
for a movement of two halves of a perfect crystal with respect to each 
other.  
 
Second and more important is the perception that the mutual motion 
along the interface would be impossible, because the interface would not 
be strictly planar (Fig.\ \ref{waves}). The interface deviation from 
planarity by just $\lesssim 10$ atomic planes is likely
enough to ensure that it is not weakened in comparison with the weaker 
of the two touching crystals.

\section{Epitaxial growth of Coulomb crystals and their elastic 
properties}
\subsection{Epitaxial freezing and crystal stretching in neutron star 
crusts and white dwarf cores}
\label{EPI}
We shall now turn to the problem of solidification in a 
compact star. Consider gradual freezing of a one-component crystal 
(100\% of ions have charge $Z e$) and suppose that the freezing 
direction is vertical, while the surfaces of simultaneous freezing are 
horizontal. Previous, already frozen layers create a set of potential 
maxima and minima characteristic of a particular crystallographic plane. 
By contrast, liquid, on average, is uniform, neutral, and approximately 
the same near any crystal plane.

\begin{figure*}
\begin{center}
\leavevmode
\includegraphics[bb=72 430 568 740, width=170mm]{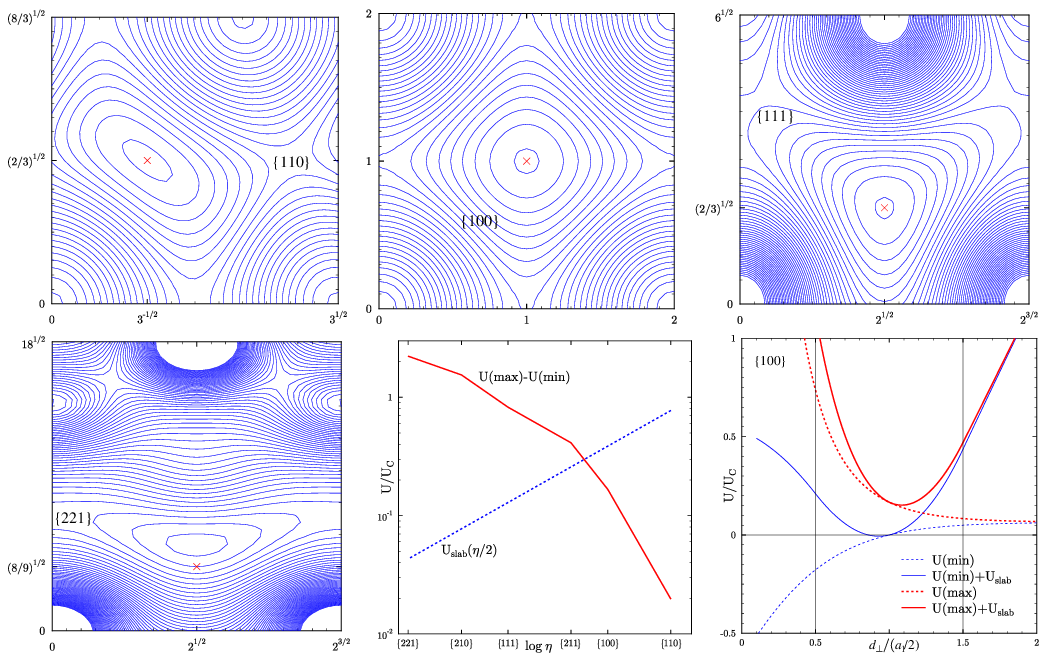}
\end{center}
\vspace{-0.2cm}
\caption[ ]{Panels a--d: potential relief produced by a semi-infinite 
bcc Coulomb crystal split along $\{110\}$, $\{100\}$, $\{111\}$, or 
$\{221\}$ crystallographic plane, respectively, vs. lateral coordinate 
$\bm{d}_\parallel$ (in units of $a_{\rm l}/2$). Vertical distance from 
the top plane $|\bm{d}_\perp|=|\bm{a}_{3\perp}|=\eta$.
Crosses indicate ion position on the next (newly forming) plane in a perfect 3D lattice. 
Potential minima are at or near the centres of the graphs, maxima are 
at the corners and/or at the graphs' upper edges. Panel e: the 
difference between the maxima and minima (solid) and $U_{\rm slab}(\eta/2)$ 
(dashed) vs. $\log{\eta}$. Panel f: potential of the $\{100\}$ 
half-lattice (dashed) and combined potential of the 
half-lattice and the electron slab (solid) 
vs. $|\bm{d}_\perp|$ with lateral coordinate, corresponding 
to a minimum (thin) or to a maximum (thick) in panel b. Vertical lines 
are the electron slab boundaries. Potential zero level is taken 
mid-slab, i.e. at $|\bm{d}_\perp|=a_{\rm l}/2$, at the lateral position 
of the minimum.}
\label{rlf}
\end{figure*}

Examples of the potential relief for various crystallographic planes 
are shown in Figs.\ \ref{rlf}a--d under the assumption that the frozen part 
of the crystal has a well-defined top plane and that crystal ions are 
located precisely at the nodes of the bcc lattice. Crystallographic 
notation is 
explained in detail in section \ref{BSSM} and Table \ref{latparam}. 
The contours in Figs.\ \ref{rlf}a--d are calculated using equation 
(\ref{U}) and plotted on 
the plane of the 2D vector $\bm{d}_\parallel$ (in units of $a_{\rm l}/2$). 
Potential minima are at or near the centres, whereas potential 
maxima are at the corners and/or at the upper edges of the plots. 
The difference between maximum and minimum potential (in units of 
$U_{\rm C}$) is shown by the solid red curve in Fig.\ \ref{rlf}e  
as a function of the respective 3D lattice interplane spacing 
$|\bm{a}_{3\perp}|=\eta$ in logarithmic scale.  

The periodic potential in Figs.\ \ref{rlf}a--d is a monotonous
function of the distance $|\bm{d}_\perp|$ from the last frozen 
plane (e.g. dashed curves in Fig.\ \ref{rlf}f).
However, if one takes 
into account the potential of the electron slab associated with the 
newly forming ion layer (this contribution is independent of the 
lateral coordinate), the combined potential acquires a minimum as a 
function of the vertical position very close to the middle of the slab 
(cf. solid curves in Fig.\ \ref{rlf}f). More precisely, ion deviation 
from the mid-slab position by a length $\delta$ results in an 
addition of $U_{\rm slab}(\delta)$ to the ion potential energy, where 
\begin{equation}
         U_{\rm slab}(\delta) = 2 \upi Z e^2 \sigma \times \begin{cases}
         \delta^2/\eta, & \delta<\eta/2 \\
         \delta-\eta/4, & \delta>\eta/2~. \end{cases}
\end{equation}
For that reason, in Figs.\ \ref{rlf}a--d, the height above the last 
frozen plane is taken equal to the respective interplane distance 
(or the electron slab thickness, 
$|\bm{d}_\perp|=|\bm{a}_{3\perp}|=\eta$).
$U_{\rm slab}$ at $\delta=\eta/2$ 
is shown by the dashed blue curve in Fig.\ \ref{rlf}e. For comparison,
let us note that kinetic energy of a typical ion at freezing is 
$\sim (3/2) U_{\rm C}/175$ \citep[e.g.][]{HPY07}, which is much less 
than the typical energy scale of potentials shown in Fig.\ \ref{rlf}. 

Thus, lateral ion positions in a newly freezing layer are determined by 
the ions of the previous layers in analogy with liquid-phase epitaxial 
crystal growth or solidification from melt, the processes well-known
in solid-state physics and semiconductor industry 
\citep[e.g.][and references therein]{BCF51,C54,B55,JC56,P56,JUH67,S82,
BS84,K91,S00}. Even though crystallization kinetics is a very 
complicated process \citep[e.g.][]{D59}, we believe, 
that due to intrinsically less hindered nature of 2D nucleation as 
opposed to 3D nucleation; 
due to lack of any ion-orientation dependence (typical 
of covalent bonds); due to long-range almost pure Coulomb forces, 
extending over a major portion of the elementary cell and capturing 
unbound ions (cf. especially Figs.\ \ref{rlf}a--c); due to a strong 
pull on ions to settle at a correct height above the already 
crystallized surface (related to charge 
neutrality); and due to extremely slow, with plenty of time for anneal, 
near-equilibrium nature of freezing in compact stars, it should 
be even easier to accomplish epitaxial growth of 
Coulomb crystals in dense matter than, say, silicon in Earth 
laboratories.

For growth perpendicular to crystallographic surfaces with low Miller 
indices (Figs.\ \ref{rlf}a--c), the lateral positions of potential 
minima produced by already frozen half-space are exactly the same as 
the lateral positions of the ions of the new layer in the infinite 3D 
lattice (crosses in Figs.\ \ref{rlf}a--d). However, crystals 
can grow perpendicular to planes of lower symmetry \citep[cf.][]{EC55,
WS62,BS84} and, in such cases, it is not necessarily so (e.g. Fig.\ \ref{rlf}d). 
It is likely, that the 
absence of the upper crystallized half-space then results in a built-in 
deformation near the interface non-uniform in the vertical direction
\citep[a deformation of this kind was discussed e.g. by][]{C59}. 
Horizontal deviations (the same for all ions) from the 3D lattice 
positions decrease with 
depth increase. As new layers are added to the top, the deformation 
moves upwards preserving self-similarity. Note, that the red curve in 
Fig.\ \ref{rlf}e displays the difference between maxima and minima as 
calculated for a perfect half-lattice (blue contours) and does not take 
this deformation, likely affecting several frozen planes at the top, 
into account.   

As we have just seen, lateral positions of ions, being added to a 
growing crystal, are fixed by previous layers. Vertical positions of the 
ions, on the other hand, are determined by charge neutrality. Thus, in 
a freezing star, interplane distances gradually increase, tracing 
$n_{\rm e}$ decrease, associated with pressure decrease as one moves 
away from the stellar center. The typical length-scale of 
$n_{\rm e}$ variation is of the order of the pressure scale height, 
$h_P=P/\nabla P$ (more on these scales in section \ref{SNGL}). 
This results in a formation of stretched (more precisely, elongated) 
ion crystals in the terminology of \citet[hereafter Paper I]{BK17}. For 
self-consistency, 
we note that, qualitatively, the general picture of Fig.\ \ref{rlf} is 
preserved for stretched matter.

Infinite 3D elongated Coulomb crystals develop unstable phonon modes, 
if the elongation exceeds a critical value (Paper I). Thus, 
overstretched crystal layers lose stability and get 
destructed as soon as their volume properties begin dominating. Since
these layers have already cooled below melting, they can then freeze 
anew, somewhat above the destruction edge, into a stretch-free cubic 
structure. Above this new cubic seed, the process of freezing with 
gradual stretching will repeat itself. Below, the cubic solid will form 
an interface with stretched (but not overstretched) layers from the 
previous step, which have survived the destruction.

It is difficult to predict the specific structure of the interface.  
Presumably, the cubic and stretched crystals will be incommensurate, 
which means zero interaction energy. However, as explained in the end 
of section \ref{NSF}, their mutual motion along the interface will be 
likely impeded by interface roughness. The shear strength of the 
interface will be about the same as that of the stretched layers below 
the interface (cf.\ section \ref{BSSM}). 

A similar process takes place when the freezing front crosses a shell 
boundary of the ``onion'' structure, from $Z_2$- to $Z_1$-enriched 
region with $Z_1<Z_2$. Let us suppose that the composition changes 
from 100\% $Z_2$ to 100\% $Z_1$, and the ratio $Z_2/Z_1$ is such that a 
crystal-like structure with a regular arrangement of lattice nodes 
can exist for any intermediate composition. It is 
natural to assume that the mixed layer is thin, so that, to zeroth 
order, there is no change of $n_{\rm e}$, which is determined by the 
pressure profile of the star. Then again, already frozen layers anchor 
the lateral positions of ions of the newly forming layers. There is, 
however, a drop in the avergage ion charge number $\langle Z \rangle$ 
from $Z_2$ to $Z_1$. Since $n_{\rm e}$ is fixed, this necessitates 
a reduction of vertical lattice spacing by a factor 
$\langle Z \rangle/Z_2$ and thus, a formation of a contracted 
crystal (Paper I) with progressively increasing degree of 
contraction.

Infinite 3D contracted Coulomb crystals are unstable for contraction 
factors as high as 0.9--0.95 (strictly speaking, this result is
obtained in Paper I only for one-component crystals). Thus, one can 
expect breaking of an overcontracted crystal, its refreezing into a 
stress-free cubic configuration, and the process repeating itself 
several times before the mixed region is fully crossed. One can expect 
a formation of several interfaces of incommensurate crystals, whose 
strength will be determined by the weaker of them, i.e., 
presumably, by a contracted crystal, with the contraction factor 
slightly above the critical value and the lowest $\langle Z \rangle$
available.   

In order to expose as vividly as possible the idea of crystal 
stretching in solidifying dense stellar matter in response to electron
density or $\langle Z \rangle$ decrease, we have assumed the simplest
possible layer-by-layer mode of the epitaxial growth. We recognize
that other growth modes are possible, for instance, edgewise growth
of certain special, in particular, close-packed planes, 
having a non-zero angle with the average crystal-liquid boundary 
\citep[e.g.][]{B55,RT57,T58,BS84}. However, the net macroscopic result 
of such growth is still a single 3D crystal, growing perpendicular to some 
general plane, constituting the average interface, which, in the case of NS and WD, must be accompanied by 
stretching to respect charge neutrality. 
Solidification of 
solutions adds yet another layer of complexity to the epitaxial process 
\citep[e.g.][]{T+53,BPS53,TB61,MS64,T68}.

To summarize, we propose, that, in contrast to the standard
picture, which is based on comparison of energies of various crystal 
structures but neglects their growth kinetics,
NS crusts and WD cores, for the most part, 
are made of stretched (elongated and contracted) crystals rather than 
of the cubic ones.  This has several astrophysical implications, the 
most obvious being for elastic properties of matter.

\subsection{Breaking strain of dense stretched matter} 
\label{BSSM}
Let us analyze in some detail the strength of the stretched crystals
with respect to shear deformations in planes orthogonal to the stretch.
This seems to be relevant for NS and WD physics,
where the stretch is aligned with the gravity, whereas the shear, 
caused, for instance, by magnetic field evolution, is horizontal and 
does not perturb hydrostatic equilibrium.

\begin{figure}
\begin{center}
\leavevmode
\includegraphics[bb=7 12 685 557, width=84mm]{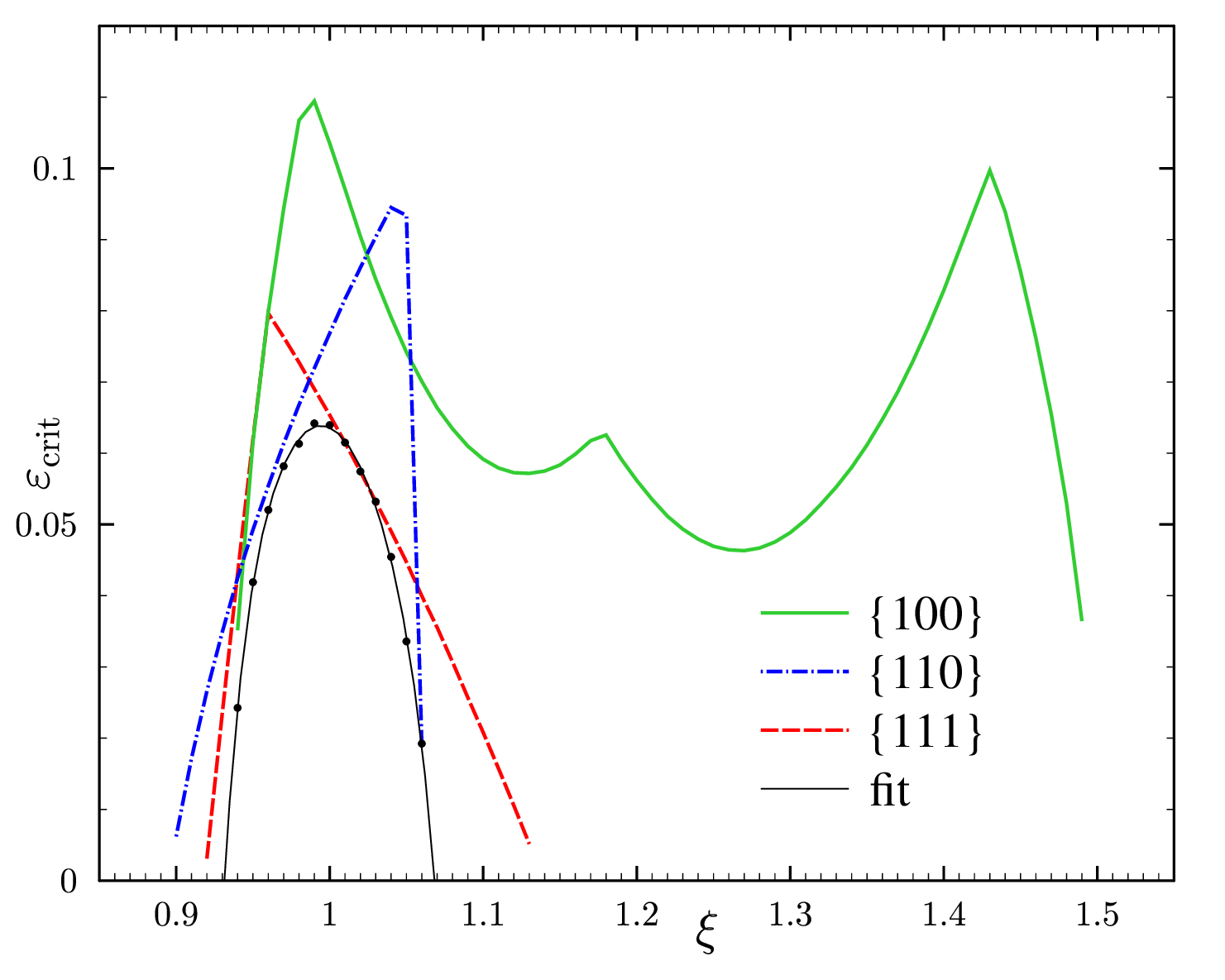}
\end{center}
\vspace{-0.2cm}
\caption[ ]{Breaking strain for transverse shear of stretched 
bcc Coulomb crystal vs. the stretch factor $\xi$.}
\label{break}
\end{figure}

In order to find breaking strain, we looked for unstable phonon 
modes near\footnote{We scanned non-equivalent spherical angles of 
a phonon wave vector assuming its length equal to 
$(1/50) 2 \upi/a_{\rm l}$, where $a_{\rm l}$, for a deformed system, is 
defined by $n_{\rm i} a_{\rm l}^3 \xi = 2$} the Brillouin zone center 
(i.e. modes with imaginary 
frequencies, see Paper I for details) for crystals, that were stretched 
by a factor $\xi$ and then sheared in the perpendicular plane. 
In practice, we have selected two basis lattice vectors 1 and 2 in the 
plane perpendicular to the stretch direction. The third basis vector 
goes out of the plane. For stretch, its vertical component is 
multiplied by $\xi$, after which, for shear, its horizontal cartesian 
components are augmented by $\theta_1 = \theta \cos{\chi}$ (along the 
basis vector 1) and $\theta_2 = \theta \sin{\chi}$. Cartesian 
components of these basis vectors and reciprocal lattice basis vectors 
as well as some other lattice parameters are summarized in Table 
\ref{latparam}.
 
With the standard strain tensor definition 
$u_{ij} = 0.5 (\partial u_i/\partial r_j+\partial u_j/\partial r_i)$,
where $u_i$ is the displacement, our shear deformation corresponds to
the appearance of non-zero components 
$u_{\theta \eta}=u_{\eta \theta} \equiv \varepsilon/2 = \theta/(2\eta)$.
At each $\xi$, we scanned the azimuthal angle $\chi$ searching for the 
minimum over $\chi$ breaking strain, $\varepsilon_{\rm crit}$.   

In Fig.\ \ref{break}, by curves of different types and colours, we 
show $\varepsilon_{\rm crit}$ for three high-symmetry 
stretch directions (associated with transverse high-symmetry planes). 
These are stretches along the cube diagonal\footnote{Stretch direction 
(A) from Paper I} of the bcc lattice (i.e. towards one of the 
nearest neighbours), along the cube edge\footnote{Stretch direction (B) 
from Paper I}, and along the cube face diagonal. In 
\citet[figs.\ 1 and 2]{BC18}, 
these stretches were defined by spherical angles 
$\theta=\tan^{-1}{\sqrt{2}}$, $\phi=\upi/4$, $\theta=0$, $\phi$ 
arbitrary, and $\theta=\upi/4$, $\phi=0$, respectively. In the 
unstretched system, the associated transverse planes cross
the axes of the cartesian reference frame aligned with the cube (with
an ion at the origin) at coordinates\footnote{in units of $a_{\rm l}$}
$(x,y,z)=(1,1,1)$, $(1,\infty,\infty)$, and $(1,1,\infty)$, 
respectively. In crystallography, these planes are members of the
equivalency classes denoted by Miller indices $\{111\}$, $\{100\}$,
and $\{110\}$.

\begin{table*}
\begin{center}
\begin{tabular}{ccccccc}
\hline
\hline
   & direct lattice basis (units $a_{\rm l}/2$) 
   & reciprocal lattice basis (units $2 \upi /a_{\rm l}$) 
   & Miller index $\vert$ $2\eta/a_{\rm l}$ $\vert$ $N_{\rm eqv}$ \\
\hline
           & $(\sqrt{8},0,0)$ & $(1/\sqrt{2},-1/\sqrt{6},
           -\sqrt{3/2}(\sqrt{2}+\theta_1)/\xi
           +(\sqrt{2/3}+\theta_2)/(\sqrt{2}\xi))$ 
           &  $\{ 111 \}$ \\
 cube-diagonal & $(\sqrt{2},\sqrt{6},0)$ & $(0,\sqrt{2/3},
           -\sqrt{2}(\sqrt{2/3}+\theta_2)/\xi)$ & $\xi/\sqrt{3}$  \\
           & $(\sqrt{2}+\theta_1,\sqrt{2/3}+\theta_2,\xi/\sqrt{3})$ 
           & $(0,0,\sqrt{12}/\xi)$ &  8 \\
\hline
           & $(2,0,0)$ & $(1,0,-(1+\theta_1)/\xi)$ 
           &  $\{ 100 \}$ \\
 cube-edge & $(0,2,0)$ & $(0,1,-(1+\theta_2)/\xi)$ & $\xi$ \\
           & $(1+\theta_1,1+\theta_2,\xi)$ & $(0,0,2/\xi)$ &  6 \\
\hline
           & $(\sqrt{3},0,0)$ & $(2/\sqrt{3},1/\sqrt{6},
           -(\sqrt{6}+\sqrt{8}\theta_1+\theta_2)/(\sqrt{12}\xi))$ 
           &  $\{ 110 \}$ \\
 face-diagonal & $(-1/\sqrt{3},\sqrt{8/3},0)$ & $(0,\sqrt{3/2},
           -\sqrt{3}(\sqrt{2/3}+\theta_2)/(2\xi))$ & $\sqrt{2}\xi$ \\
           & $(1/\sqrt{3}+\theta_1,\sqrt{2/3}+\theta_2,\sqrt{2}\xi)$ 
           & $(0,0,\sqrt{2}/\xi)$ &  12 \\
\hline
           & $(\sqrt{20},0,0)$ & $(1/\sqrt{5},0,
           -(3/\sqrt{5}+\theta_1)/\xi)$ &  $\{ 210 \}$ \\
 aux 1     & $(0,2,0)$ & $(0,1,-\sqrt{5}(1+\theta_2)/\xi)$ 
           & $\xi/\sqrt{5}$ \\
           & $(3/\sqrt{5}+\theta_1,1+\theta_2,\xi/\sqrt{5})$ 
           & $(0,0,\sqrt{20}/\xi)$ &  24 \\
\hline
           & $(\sqrt{8},0,0)$ & $(1/\sqrt{2},0,
           -\sqrt{3}(\sqrt{2}+\theta_1)/(2\xi))$ 
           &  $\{ 211 \}$ \\
 aux 2     & $(0,\sqrt{3},0)$ & $(0,2/\sqrt{3},
           -\sqrt{2}(1/\sqrt{3}+\theta_2)/\xi)$ 
           & $\xi\sqrt{2/3}$ \\
           & $(\sqrt{2}+\theta_1,1/\sqrt{3}+\theta_2,\xi\sqrt{2/3})$ 
           & $(0,0,\sqrt{6}/\xi)$ &  24 \\
\hline
           & $(\sqrt{8},0,0)$ & $(1/\sqrt{2},-1/\sqrt{18},
           -3(7/9+\theta_1/\sqrt{2}-\theta_2/\sqrt{18})/\xi)$ 
           &  $\{ 221 \}$ \\
 aux 3     & $(\sqrt{2},\sqrt{18},0)$ & $(0,\sqrt{2}/3,
           -\sqrt{2}(\sqrt{8}/3+\theta_2)/\xi)$ 
           & $\xi/3$ \\
           & $(\sqrt{2}+\theta_1,\sqrt{8}/3+\theta_2,\xi/3)$ 
           & $(0,0,6/\xi)$ &  24 \\
\hline
           & $(\sqrt{20},0,0)$ & $(1/\sqrt{5},-\sqrt{3/35},
           \sqrt{21}(2/7-\theta_1/\sqrt{5}
           +\sqrt{3/35}\theta_2)/\xi)$ &  $\{ 421 \}$ \\
 aux 4     & $(6/\sqrt{5},\sqrt{84/5},0)$ & $(0,\sqrt{5/21},
           -\sqrt{5}(17/\sqrt{105}+\theta_2)/\xi)$ 
           & $\xi/\sqrt{21}$ \\
           & $(1/\sqrt{5}+\theta_1,17/\sqrt{105}+\theta_2,
           \xi/\sqrt{21})$ & $(0,0,\sqrt{84}/\xi)$ &  48 \\
\hline
\end{tabular}
\end{center}
\caption{Lattice parameters}
\label{latparam}
\end{table*}

One observes striking differences in breaking strain behaviour with 
$\xi$ for different stretch directions. A very significant elongation
is possible along the cube edge (Paper I). At $\xi=\sqrt{2}$, 
the lattice, that was originally bcc, turns into face-centred cubic 
(fcc). 
For stretches 
along the cube diagonal, the breaking strain drops abruptly for 
contractions but gradually for elongations. For stretches along the 
face diagonal, the situation is exactly opposite.

It is not known what is the actual structure of NS crust and 
what is the proper way of deducing crust properties from those
of perfect crystallites. 
A plausible model ({\it polycrystalline} model) may be to assume 
that, in a horizontal layer, there are crystallites stretched vertically 
by approximately the same factor $\xi$ but oriented more or less 
randomly. Experiments on bicrystals show that crystals with 
different liquid-solid interface orientations can grow side by side
\citep[e.g.][fig.\ 3]{RT57}. 
Then, under a shearing deformation, the crystallites 
will have the same strain to maintain continuity, and the crystallite 
with the {\it minimum breaking strain} will fail first. This may result in a 
stability loss and failure of neighbouring 
crystallites\footnote{Another alternative is to assume that at any depth there are 
crystallites with all possible $\xi$. Then one would also have to
minimize breaking strain over $\xi$, which would mean that at any depth
the minimum breaking strain would be arbitrarily close to zero
(cf.\ section \ref{SNGL}).}.

In order to have a more representative breaking strain minimization,
we have additionally considered 4 stretch directions of lower symmetry.
Lattice properties for these cases, denoted aux 1--4, are also 
collected in Table \ref{latparam}. We note that the 7 considered cases 
of crystal growth differ, among other things, by interplane spacing 
$\eta$, which varies (for $\xi=1$) from $a_{\rm l}/\sqrt{84}$ for 
auxiliary case 4 to $a_{\rm l}/\sqrt{2}$ for the face-diagonal stretch
(cf.\ last column of Table \ref{latparam}). 

The numerical data for the minimum breaking strain, obtained by minimizing over 
these 7 stretch directions, is shown in Fig.\ \ref{break} by dots. 
To rms accuracy better than 0.1\%, the data can be fitted as  
\begin{eqnarray}
      \varepsilon_{\rm crit}^{\rm min} &=& 0.0639-20.1\,(\xi^{2/3}-0.99625)^2
\nonumber \\
                &-&2.33\mathrm{e}6 \,(\xi^{2/3}-1.00145)^6~.
\label{epsfit}
\end{eqnarray}

Breaking strain of 0.12--0.14 for shear of an ideal bcc crystal at 
$\xi=1$ has been obtained by \citet[fig.\ 1]{HK09} in MD simulations. 
This value has been reduced to 0.1 to account for various 
imperfections and is currently used in applications as the breaking 
strain for shear deformations \citep[see also discussion in][]{BC18}. 
Minimum breaking strain for shear of the ideal unstretched bcc 
crystal obtained in the present work [Fig.\ \ref{break} and equation 
(\ref{epsfit})] is $\approx 0.064$, which is $\sim 2$ times lower 
than the range quoted above. Reducing it in the same way, we arrive at 
the updated value of $\sim 0.05$ as the breaking shear strain 
of an imperfect unstretched bcc crystal.

\subsubsection{Preferred orientation of growth and macrocrystallites}
\label{SNGL}
Liquid-phase epitaxy is a near-equilibrium process, which produces 
crystal layers of extremely high quality. Also of interest is the 
Czochralski crystal growth technique \citep[][]{C1918,TL50,B51,H87,S00}, 
which nowadays yields 
ideal crystals of $\sim 2$ m size. We note that these methods are 
subject to severe limitations posed by finite sizes of 
the apparatus and associated with them nonuniformities, thermal 
gradients, and stresses \citep[e.g.][]{D59}. It seems then, that in a 
near-equilibrium 
crystallization with extremely uniform temperature and composition 
distributions and plenty of anneal time, NS crusts and especially WD 
cores (which literally take eons to freeze) have a much better chance 
of forming large-scale near perfect crystallites
\citep[this possibility has been mentioned earlier by][]{HK09}. 

On Earth, natural single 
crystals as large as $\sim 18$ m (and possibly $\sim 50$ m) have been 
found \citep[][]{R81}. The abstract of this paper begins with a 
remarkable statement: ``No upper limit on the size of crystals is to be
expected \ldots \,'' The occurrence of so huge natural crystals 
indicates, at least, that there are robust self-consistent mechanisms 
of seeding their growth. 

In an illuminating experiment of \citet[figs.\ 6 and 8]{RT57}, purified 
lead was melted and poured into a mold designed for unidirectional 
(upward) freezing. Subsequent analysis of the ingot had shown that 
there were no crystallites other than those, which nucleated on the 
bottom boundary, and that the ingot bottom surface had 10 times as many 
crystallites as the top one. This means that 90\% of the original 
crystallites have been crowded out by the surviving ones. A possible 
theoretical model of such crowding out has been proposed by 
\citet[esp.\ fig.\ 8]{T57}.
The preferred orientation of crystallite growth was close to 
perpendicular to $\{111\}$ planes. With addition of a sufficient 
quantity of silver impurities, the preferred orientation switched to 
that perpendicular to $\{100\}$ planes. However, the conditions of 
these experiments were clearly far from equilibrium. 

According to Bravais's rule, in equilibrium, crystals tend to grow 
towards a shape bounded by the slowest growing planes 
\citep[e.g.][]{TB61}. The fastest growing surfaces grow themselves
out \citep[e.g.][]{WS62}. The slowest growing planes are the 
close-packed ones. These are $\{110\}$ planes for bcc crystals and 
$\{111\}$ planes for fcc. The rule is supported by prominence of 
$\{111\}$ plane growth in various experiments on Earth oftentimes 
conducted on fcc materials. Should we then expect that the entire 
crystallization front in a compact star grows perpendicular to 
$\{110\}$ planes of the stretched bcc lattice or close to these 
directions? This is a distinct possibility\footnote{Another possibility, 
very similar energetically and consistent with the Bravais's rule, is 
obviously the front growth perpendicular to $\{111\}$ planes of the 
stretched fcc lattice.}.
However, we note, that in a rare experiment on a bcc material sodium
\citep[][fig.\ 1]{WS62}, no clear orientational preference has been
observed with a slight tendency to directions perpendicular to 
$\{111\}$.  

In the case of macroscopic crystallite formation ({\it macrocrystallite} model), particular properties 
of ideal crystallites may have relevance for astrophysical phenomena.
For instance, if we assume $\{ 111 \}$ growth of a bcc lattice (with 
$\xi_{\rm crit} \approx 1.13$), the critical elongation is achieved 
over the length of $\sim 0.13 n_{\rm e}/\nabla n_{\rm e}$, where  
\begin{equation}
   \frac{n_{\rm e}}{\nabla n_{\rm e}} = \frac{4}{3} h_P =
   \frac{4P_{\rm e}}{3\rho g} = \frac{4}{3} 
   \left( \frac{3Z}{A}\right)^{4/3} \left(\frac{3\mathrm{e}14}{g}
   \right) \, \rho^{1/3} ~~{\rm cm}, 
\end{equation}
and we have assumed that the pressure is dominated by 
ultrarelativistic electron gas. Furthermore, $\rho$ is the mass density 
in units of g/cc, $g$ is the gravitational acceleration in cm/s$^{2}$, 
$P_{\rm e}$ is the electron pressure, and $A$ is the ion mass number. 
At density $10^9$ g/cc, it takes $\lesssim 2$ m to grow to a critical 
height\footnote{We therefore do not expect $\sim 10$ m tall 
crystallites in the outer layers of NS mentioned in a slightly
different context by \citet{C+18}, but 
lateral sizes of this magnitude or more do not seem impossible.}.   

At $\xi$ not too close to $\xi_{\rm crit}$, the elastic properties of 
matter differ quantitatively but not qualitatively from the 
conventional results. We see however, that 
the descending portion of the dashed red curve in Fig.\ \ref{break} is 
approximately linear from $\xi=1$ to $\xi_{\rm crit}$. This indicates 
the presence, upon freezing, of layers (with 
$\xi \lesssim \xi_{\rm crit}$) whose breaking strain for shear is
much lower than at $\xi=1$. Specifically, there is a layer of 
$\lesssim 20$ cm thick, in which the breaking strain is 10 times lower, and 
a layer of $\lesssim 2$ cm thick, in which the breaking strain is 100 
times lower, than for the bulk of the crystallite. 
Let us clarify that, even though these layers are substantially weaker
than the unstretched material, they are stable and, by any 
conventional measure, extremely strong. They would not break 
spontaneously, but require for that a huge stress, which, however, 
is 1-2 orders of magnitude lower than at $\xi \approx 1$. 
If, under the action of magnetic or any other such stresses late in a 
NS evolution, these layers 
break and then refreeze into a stress-free cubic configuration, they 
become 10 or 100 times stronger and thus, much harder to break at a 
later time. 

If, on the other hand, the growth is of the face-diagonal type with
$\xi_{\rm crit} \approx 1.06$, the maximum crystallite height at the 
same density is $\sim 0.8$ m. In the course of stretching, the 
crystal becomes some 20\% stronger and then abruptly loses stability 
(cf. dot-dashed blue curve in Fig.\ \ref{break}). Due to extremely 
large derivative $|{\rm d}\varepsilon_{\rm crit}/{\rm d}\xi|$ near 
$\xi_{\rm crit}$ for elongations, weak layers, if any, are much thinner.

\subsection{Effective shear modulus}
\label{ESM}
\begin{figure}
\begin{center}
\leavevmode
\includegraphics[bb=6 15 695 557, width=84mm]{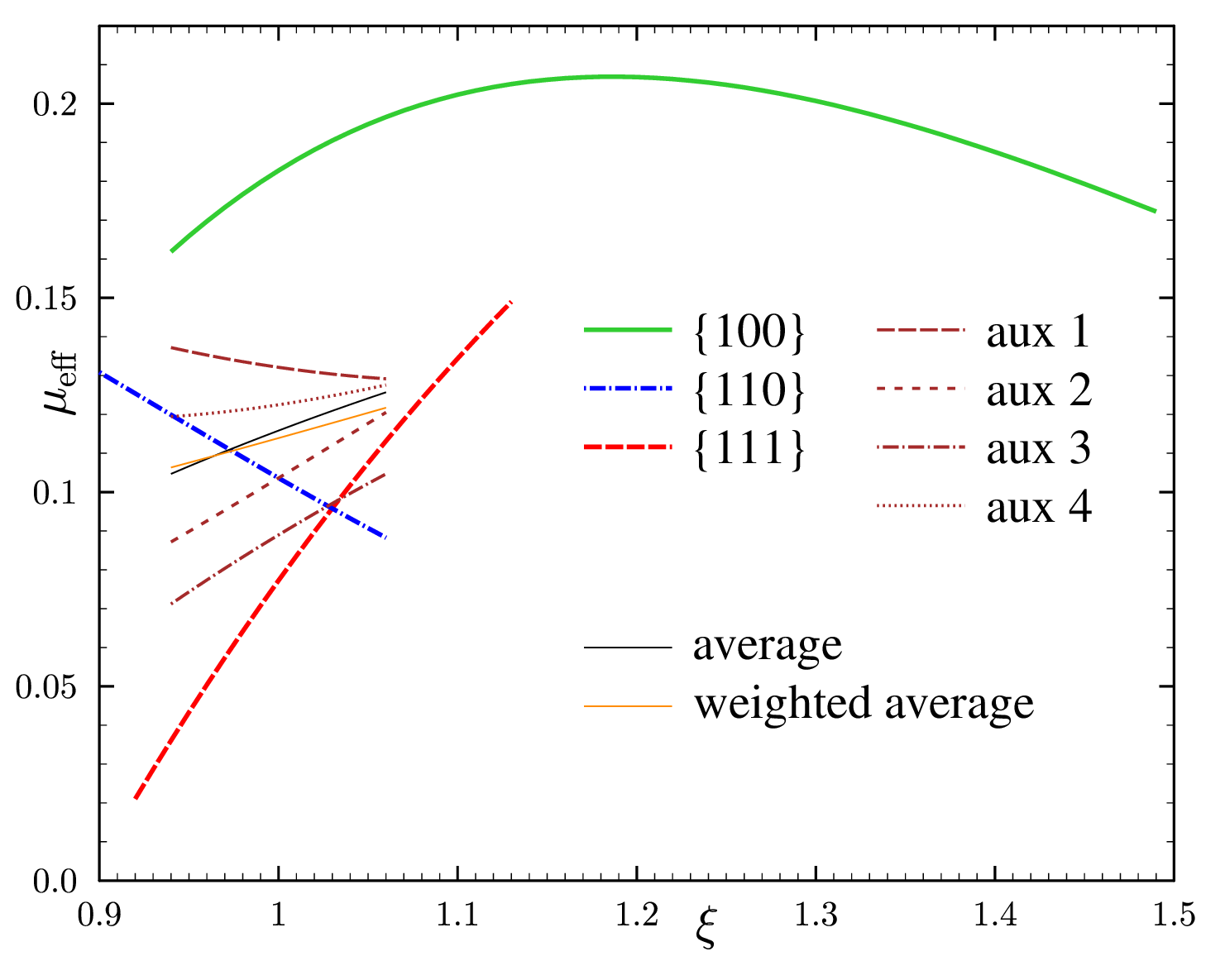}
\end{center}
\vspace{-0.4cm}
\caption[ ]{Effective shear modulus of stretched bcc Coulomb crystal in
units of $n_{\rm i}U_{\rm C}$.}
\label{muaver}
\end{figure}

Another quantity of interest for applications is the effective shear 
modulus 
$\mu_{\rm eff}(\xi)$. In order to calculate it at arbitrary $\xi$, one 
could 
apply an infinitesimal shearing deformation to a crystal stretched in a 
particular direction, evaluate the 
energy difference $\delta U$, and then derive shear modulus $\mu$ for 
this stretch orientation from the formula:
\begin{equation}
        \frac{\delta U}{V} = 2 \mu u^2_{\theta \eta}~,
\end{equation}
where $V$ is the volume. Given a particular stretch direction, firstly, 
we have found $\mu_{\rm eff}$ averaged over the shearing deformation azimuthal 
angle $\chi$. This quantity (in units of $n_{\rm i} U_{\rm C}$, 
assuming single ion type with charge $Ze$) is 
shown in Fig.\ \ref{muaver} as a function of $\xi$ for 7 different 
stretch orientations. The variation in some cases is remarkable. For the
cube-diagonal stretch (dashed red), $\mu_{\rm eff}^{\{111\}}$ varies from 0.02 at $\xi=0.92$ to 0.15 at 
$\xi=1.13$. For cube-edge stretch (solid green), we reproduce classic 
results for $S_{1212}$ for bcc (0.1827) and fcc (0.1852) lattices
\citep[][]{F36}.  

Since strain for all crystallites is assumed to be the same, we can 
further average $\mu_{\rm eff}$ over 7 stretch orientations. This results in a 
slowly varying thin solid black curve in Fig.\ \ref{muaver}. 
This kind of averaging neglects the fact that some planes are nominally 
less numerous than others,
for instance, there are only 6 original cubic seed orientations, that 
produce growth perpendicular to a $\{ 100 \}$ plane, whereas there 
are 48 equivalent $\{ 421 \}$ planes (aux 4). Qualitatively, such 
simplistic averaging may reflect
the truth if the growth of low-index planes is, for whatever reason,
preferred, or if they effectively represent larger solid angles. 
Optionally, we can treat all planes on the same footing and include
the number of equivalent planes $N_{\rm eqv}$ into the average shear 
modulus calculation\footnote{$N_{\rm eqv}$ is given in the last column 
of Table \ref{latparam}.}. Such weighted average results in the slightly 
different thin solid orange curve in Fig.\ \ref{muaver}. At $\xi=1$, 
these approaches 
yield $\mu_{\rm eff}(1) \approx 0.116$ and 0.114, respectively, which, 
in both cases, is within 5\% of the effective shear 
modulus $\mu_{\rm eff}^{\rm OI}=0.1194$ introduced by \citet{OI90}. 

At arbitrary $\xi$, various curves in Fig.\ \ref{muaver} 
can be fitted to rms accuracy better than 0.01\% 
(better than 0.1\% for $\{100\}$) as
\begin{eqnarray}
        \mu_{\rm eff}^{\{100\}} &=& 1.8 - 1.157 \, \xi^{1/2} - 0.46 \, \xi^{-1.89}~,  \\
        \mu_{\rm eff}^{\{110\}} &=& 0.895 - 0.79132 \, \xi^{1/3}~,  \\
        \mu_{\rm eff}^{\{111\}} &=& 0.0491+\xi^{-1/3}-0.97182 \, \xi^{-1}~,  \\
        \mu_{\rm eff}^{\rm av} &=& 2.08 - 1.06418 \, \xi^{-1/3} - 0.9 \, \xi^{1/5}~,  \\
        \mu_{\rm eff}^{\rm wav} &=& -1.94 + 0.95392 \, \xi^{-1/4}+1.1 \, \xi^{1/3}~.  
\end{eqnarray}

\subsection{Other possible effects}
\label{OPE}
In this subsection, we shall list a few other possible effects of 
crystal stretching in NS crusts and WD cores, which have not been 
studied in any detail but may be relevant for astrophysics. 

It is clear, that stretched crystals have higher ground-state 
energy, than the cubic ones. The difference between the two has been 
analyzed in some detail by \citet[figs.\ 5 and 6]{BC18}, where it was 
shown to be less than the latent heat associated 
with freezing into the stress-free cubic lattice. We thus
expect, that freezing into a stretched configuration produces less 
latent heat at crystallization than the standard value. The rest 
of the energy, in principle, may be released at an arbitrarily late 
cooling stage. In the context of WD, this may happen at an age, 
when the stellar luminosity is much lower than at crystallization, and 
thus noticeably delay WD cooling.  
 
Phonon thermodynamics of stretched crystals in general, and their heat 
capacity in particular may be subject to appreciable modifications.
This is related to the expected appearance of a low-frequency mode,
which eventually, at $\xi_{\rm crit}$, becomes unstable. This also
may be important for cooling modelling. Phonon dispersion curves and 
polarisation vectors affect electron-phonon scattering rates, which, in 
turn, determine various electron kinetic coefficients such as thermal 
and electric conductivities. If there is a preferred crystal growth 
direction, this quantities will become anisotropic. This may be 
especially pronounced for crystals stretched perpendicular to $\{100\}$ 
planes (where $\xi$ can significantly exceed 1) or stretched close to 
their breaking limits. 

Stretching affects nearest neighbour distance in a crystal.
At fixed density, for the cube-diagonal stretch, this 
distance along the stretch direction is different from that in the 
unstretched lattice by a factor of $\xi^{2/3}$. For the face-diagonal 
stretch, the nearest neighbour distance in the transverse direction 
gets multiplied by $\xi^{-1/3}$. This may be important for pycnonuclear 
reaction rates \citep[e.g.][]{Y+06}, which are very sensitive to 
subtle details such as Coulomb barrier height and tunneling length, 
and may easily vary by many orders of magnitude.

\section{Conclusion}
Based on thermodynamic and evolutionary arguments, it is well established
that NS crusts and WD cores contain adjacent regions with
different ion compositions. At low tempertures, the regions 
crystallize. Boundary layer between these crystals is 
expected to be relatively thin so that transition between them 
occurs at essentially constant electron density 
$n_\mathrm{e}$. We have derived a general expression 
for the interaction energy of such Coulomb crystals in an idealized model. Under broad 
assumptions regarding the composition and lattice structure, the 
crystals are incommensurate, and in this case, the 
interaction energy reduces identically to zero. We hypothesize that, in 
reality, for both commensurate and incommensurate crystals, their 
interaction is actually controlled by interface roughness. Consequently, 
in spite of the interface existence, the combined system behaves 
as a single crystal, and its breaking properties are determined by those of 
the weaker part.       

Using formulae obtained on the previous step, we have revisited 
near-equilibrium bottom-up crystallization in NS crusts and WD cores.
We argue that this process is in many ways similar to liquid-phase 
epitaxial (i.e. preserving order of previous layers) crystal growth or 
crystal pulling from melt in terrestrial laboratories. Moreover, we 
believe that due to lack of electron bonds and, more generally, due to 
independence of angular orientation of newly arriving nuclei, due to 
long-range nature of Coulomb forces, and due to nearly ideally 
uniform environment of compact star interior, the epitaxial growth of 
Coulomb crystals in these astrophysical objects may be realized more 
easily than on Earth. 

If this is true, lateral ion positions are locked by already frozen 
layers, whereas vertical ion positions are governed by charge neutrality. 
Therefore, interplane spacings in growing crystals either gradually
increase, tracing $n_\mathrm{e}$ decrease, as the crystallization front
moves away from the stellar center, or decrease, tracing decrease of 
the average ion charge number, when the crystallization front crosses boundary
between the regions with different compositions. This results in a 
formation of stretched (elongated and contracted) Coulomb lattices, as 
opposed to the standard assumption of cubic lattice formation,
which is based on energetics arguments but does not take into account 
crystal growth kinetics.
Overelongated and overcontracted crystals develop unstable phonon modes, 
lose stability, and get destructed. This limits the vertical sizes of 
growing crystallites. 

We study breaking strain of stretched matter
for shear deformations perpendicular to the crystal growth direction and find
striking differences in its behaviour vs. stretch factor $\xi$ 
for different stretch orientations. For unstretched material, 
we find a shear deformation with breaking strain $\sim 2$ times lower than 
the currently known estimate (0.05 instead of 0.1 with account of 
crystal imperfections). Assuming the presence at a given depth 
of arbitrarily oriented crystallites stretched by the same factor
({\it polycrystalline} model), we 
minimize breaking shear 
strain over stretch directions at each $\xi$ and fit the resulting 
$\varepsilon_{\rm crit}^{\rm min}(\xi)$ dependence by a simple analytic formula.

Furthermore, we present some arguments in favor of macrocrystallite
formation and preferred crystal growth orientation in dense matter 
of NS crusts and WD cores ({\it macrocrystallite} model). If this is the case, specific properties of 
crystallites may be directly connected with observed astrophysical phenomena. 
For instance, for growth parpendicular to $\{111\}$ planes, we expect 
the appearance of crustal layers whose shear strength is 10--100 times
lower than for unstretched matter of the same density. If, under the action 
of magnetic or any other stresses late in an NS evolution, these layers 
break and then refreeze into a stress-free cubic configuration, they 
become 10--100 times stronger and do not break that easily at a later 
time. On the contrary, for growth parpendicular to $\{110\}$ planes, which are 
the close-packed planes of the bcc lattice, no such weak 
layers are expected. The diversity of layer sizes and strengths as well as the possibility
of significant strengthening of matter after breaking 
may be related to rich magnetar burst and outburst phenomenology 
\citep[e.g.][]{KB17} and its extension to lower-$B$ objects.

We have also calculated the dependence of the crystal shear modulus on 
$\xi$ for growth perpendicular to 3 low-index planes and 4 planes of 
lower symmetry as well as the shear modulus averaged over growth 
orientations and fitted main results by analytic expressions. Several 
other effects of crystal stretching on thermodynamics, kinetics, and 
pycnonuclear reaction rates in NS crusts and WD cores, which may be 
important for applications, have been proposed.

\section*{Acknowledgments}
This work with the exception of sections \ref{GF}--\ref{EPI} 
have been supported by the Russian Science Foundation grant 
19-12-00133-P.

\section*{Data Availability}
The data underlying this article will be shared on reasonable 
request to the author.

\end{document}